\documentclass[PRB,twocolumn,showpacs,floatfix,notitlepage,superscriptaddress]{revtex4-1}
\usepackage{graphicx,epsfig,psfrag}
\usepackage{epstopdf}
\usepackage{amsfonts}
\usepackage{appendix}
\usepackage{times}
\usepackage{graphics,dcolumn,bm,epic,eepic,float}
\usepackage{amssymb,amsmath,amsfonts,rotate,color}
\usepackage[colorlinks=true,linkcolor=blue,urlcolor=blue,citecolor=blue,anchorcolor=blue]{hyperref}
\setcitestyle{numbers,square}
\usepackage{mathptmx}
\usepackage{txfonts}
\usepackage[dvipsnames]{xcolor}

\begin{document}

\title{Detrimental effects of disorder in two-dimensional time-reversal invariant topological superconductors}
\date{\today}

\author{Mahdi Mashkoori}
\affiliation{Max Planck Institute for the Physics of Complex Systems, N\"othnitzer Str.~38, 01187 Dresden, Germany}
\affiliation{Department of Physics, K.N.~Toosi University of Technology, P.~O.~Box 15875-4416, Tehran, Iran}

\author{Fariborz Parhizgar}
\affiliation{Department of Physics and Astronomy, Uppsala University, Box 516, SE-751 20 Uppsala, Sweden}

\author{Stephan Rachel}
\affiliation{School of Physics, University of Melbourne, Parkville, VIC 3010, Australia}

\author{Annica M. Black-Schaffer}
\affiliation{Department of Physics and Astronomy, Uppsala University, Box 516, SE-751 20 Uppsala, Sweden}

\begin{abstract}
The robustness against local perturbations, as long as the symmetry of the system is preserved, is a distinctive feature of topological quantum states. Magnetic impurities and defects break time-reversal invariance and, consequently, time-reversal invariant (TRI) topological superconductors are fragile against this type of disorder. Non-magnetic impurities, however, preserve time-reversal symmetry and one naively expects a TRI topological superconductor to persist in the presence of non-magnetic impurities. In this work, we study the effect of non-magnetic disorder on a TRI topological superconductor with extended $s$-wave pairing, which can be engineered at the interface of an Fe-based superconductor and a strongly spin-orbit coupled Rashba layer. We model two different types of non-magnetic random disorder and analyze both the bulk density of states and edge state spectrum. Contrary to naive expectations, we find that the disorder strongly affects the topological phase by closing the energy gap, while trivial superconducting phases remain stable and fully gapped. The disorder phase diagram reveals a strong expansion of a nodal phase with increasing disorder. We further show the decay of the helical Majorana edge states in the topological phase and how they eventually disappear with increasing disorder. These results alter our understanding of effects of impurities and disorder on TRI topological phases and may help explain the difficulty of experimental observation of TRI topological superconductors.
\end{abstract}

\maketitle
%
%
\section{Introduction}
Much of the recent developments in condensed matter physics have focused on theoretical prediction and experimental verification of topological phases of matter\cite{WenBook,BernevigBook,Moessner2021}. Topological superconductivity constitutes a particular exciting case, thanks to the presence of exotic Majorana quasi-particles and their potential for technological applications \cite{Kitaev01,Wilczek09,Aguado2017}. With topological phases being generally robust and protected against local perturbations, such as disorder and decoherence, topological superconductors have become particularly interesting candidates for topological quantum computing \cite{nayak_non-abelian_2008,ivanov01prl268}. 

Time-reversal symmetry, alongside particle-hole and chiral symmetries, plays an essential role in classifying topological states \cite{Schnyder2008,Kitaev2009}. Several platforms for topological superconductivity have already been studied in both theory and experiment, where time-reversal symmetry is broken explicitly either by an external magnetic field or by magnetic impurities \cite{Anindya2012,Mourik2012, Churchil2013, Yazdani14, Lothman2014, Pawlak2015, Ruby17,Menard17, Wiesendanger18, Palacio18, Steinbrecher2018, Choi19, schneider_topological_2021, MashkooriDwave,Mashkoori2020}. Topological superconductivity that preserves time-reversal symmetry has also been widely predicted theoretically\,\cite{Fu2010,Wray2010,Zhang13}.

Generally, a topological phase is expected to be robust against any perturbation that does not break its classifying symmetries. In particular, a time-reversal invariant (TRI) topological phase would be expected to be robust against any perturbation that does not break time-reversal symmetry. As a consequence, several topological TRI systems have been shown to prevail in the presence of non-magnetic impurities and disorder \cite{Leijnse12,BeenakkerRMP,Mascot19,crawford-20prb174510}. Despite this strong disorder protection, TRI topological superconductivity has so far been challenging to observe experimentally. 

One promising proposal for experimental realization of a TRI topological superconductor is based on proximity-induced superconductivity from an Fe-based superconductor into a Rashba spin-orbit coupled (SOC) layer \cite{Zhang13}. Belonging to class DIII, the effective Hamiltonian for this hybrid structure in two dimensions (2D) is topologically characterized by a $\mathbb{Z}_2$ index \cite{Schnyder2008,Kitaev2009}. 
In the topological phase, a Kramers pair of helical Majorana edge states propagate along the boundary of the system. 
By simply tuning the chemical potential it is also possible to tune from the topological to a trivial phase, with an intersecting nodal phase \cite{Zhang13}. 
The effects of single magnetic and non-magnetic impurities have previously been studied in this system \cite{Mashkoori2018}. In particular, it was shown that non-magnetic impurities behave very differently in the topological and trivial phases. For any material realization and its functionality it is however more relevant to consider random disorder and its effects. 

In this work, we perform numerical calculations to investigate the effect of random non-magnetic disorder on TRI topological superconductors by studying hybrid structures composed of an Fe-based superconductor and an effective 2D material with Rashba SOC. In particular, we consider both Anderson disorder, where the chemical potential is randomly fluctuating, and concentration disorder, where randomly placed dilute but strong non-magnetic scatterers are present.  
By evaluating the bulk density of states (DOS), we find that even moderately weak non-magnetic disorder induces subgap states that quickly fill the entire energy gap in the topological phase. 

Some aspects of an disorder-induced metallic phase has previously been discussed \cite{Fulga_2012,Diez_2014} and attributed to sign-changing potential fluctuations, while we demonstrate fragility of topological phase for both small disorder and without any need for sign-changing scattering.
Furthermore, we are able to trace this disorder sensitivity of the topological phase to the existence of subgap states in the single- and few-impurity limit, although a single impurity never generates states near zero energy, such that only random bulk disorder can fully destroy the gap.
We find that the critical disorder strength that fully closes the gap is almost exactly proportional to the size of the topological gap in the clean limit. This leads to the topological phase being particularly fragile for small gaps.
We further show that this disorder behavior of the topological phase stand in sharp contrast to that of trivial conventional and extended $s$-wave superconductors, which both remain fully gapped even for strong disorder. 
Consequently, by increasing the disorder, we observe a strong expansion of the nodal region in the phase diagram, essentially only at the expense of the topological phase. 
We also study the evolution of the helical Majorana edge modes with increasing  disorder and show how they quickly delocalize into the bulk and are eventually destroyed when the bulk gap closes. These results clearly demonstrate that the naive expectation of robustness against symmetry-preserving perturbations are not accurate for TRI topological superconductors; instead they are surprisingly sensitive to non-magnetic disorder. With non-magnetic disorder being very common in most superconductors and hybrid structures, our results indicates that experimental observation of TRI topological superconductivity might be challenging.

The remainder of this work is organized as follows. In Sec.~\ref{Model}, we introduce a 2D lattice model to study the hybrid structure of an Fe-based superconductor and a Rashba SOC layer. In Sec.~\ref{Results}, we present our results where we first analyze the disorder dependence of the bulk DOS in Sec.~\ref{DOS}. Next, in Sec.~\ref{PhaseDiagram} we discuss the phase diagram in the presence of disorder, and finally in Sec.~\ref{EdgeStates} we show the effect of disorder on the topological edge states. We summarize our results in Sec. \ref{summary}.

%
\section{Model}
\label{Model}
A TRI topological superconductor can be constructed at the interface of a 2D Rashba SOC material and an Fe-based $s_\pm$-wave superconductor, as originally proposed by Zhang et al.\,\cite{Zhang13}. In this work, the authors considered a 2D square lattice with nearest-neighbor hopping $t$ and Rashba spin-orbit interaction $\lambda_R$ from the 2D Rashba SOC material and with superconducting pairing, consisting of on-site $\Delta_0$ and isotropic nearest neighbor $\Delta_1$ pairing terms, induced by the extended $s$-wave symmetry of the Fe-based superconductor. The tight-binding Hamiltonian for this system within the standard mean-field framework for superconductivity reads:
\begin{align}
{\cal H}_0=
& 
\sum_{ {\bf i},{\bf j}, \sigma}
( -\frac{1}{2}t_{{\bf ij}} c_{{\bf i}\sigma}^\dag
c_{{\bf j}\sigma}+{\Delta_{{\bf i}{\bf j}}}c_{{\bf i}\sigma}^\dag
c_{{\bf j} {\bar\sigma}}^\dag )\nonumber \\ 
& -\lambda_R \sum_{\bf i ,\eta=\pm} \eta c_{{\bf i},\uparrow}^{\dag} 
(c_{{\bf i}-\eta\hat{\bf x},\downarrow}-
ic_{{\bf i}-\eta\hat{\bf y},\downarrow}) + \textrm{H.c.}\ ,
\label{tight-binding.eq}
\end{align}
with ${\Delta_{{\bf i}{\bf j}}=\Delta_0}$ and $t_{{\bf i}{\bf j}} = \mu$ for ${\bf i}={\bf j}$, where ${\bf i} = (i_x,i_y)$ represents a site in a 2D square lattice and $\mu$ is the chemical potential.
We restrict the range of hopping and superconducting pairing to nearest-neighbors, $\langle {\bf i, j} \rangle$, with $t_{{ \bf\langle i, j \rangle}} = t$ and  $\Delta_{{ \bf\langle i, j \rangle}}=\Delta_1$. For clarity, we scale all parameters to the nearest neighbor hopping by setting  $t=1$. The combination of $\Delta_0$ and $\Delta_1$ results in an extended $s$-wave pairing symmetry that retains the symmetry of the  normal-state Hamiltonian but changes sign between the $\Gamma$ and $M$ points of the first Brillouin zone. The Rashba SOC also splits the normal-state Fermi surface, such that by adjusting the chemical potential one can make the inner and outer Fermi surfaces experience pairing gaps with either same or opposite signs. This has direct consequences for the topology of the system.
It is noteworthy that in Hamiltonian Eq.~(\ref{tight-binding.eq}), no scattering takes place between the two helical Fermi surfaces.  Furthermore, there is no interband scattering due to impurities, as the topological features protected by symmetry prevent any such interband scattering.
Time-reversal symmetry is preserved for the Hamiltonian in Eq.~\eqref{tight-binding.eq}, which means it belongs to class DIII. For this class in 2D the topological classification is through a $\mathbb{Z}_2$ index and not a Chern number \cite{Schnyder2008,Kitaev2009}. The $\mathbb{Z}_2$ invariant is set by the relative sign of the superconducting gap for each of two SOC split Fermi surfaces. Opposite signs of the gap yields a topological phase with $\nu=1$ and $s_\pm$-wave superconducting pairing, while the same sign gives a trivial phase with $\nu=0$ and $s_{++}$-wave pairing \cite{Qi2010,Teo2010}. The topological phase is known to have a Kramers pair of 1D helical Majorana states at any boundary toward a trivial region \cite{Zhang13}.

\begin{figure}[t]
\centering
\includegraphics[width=1.\columnwidth]{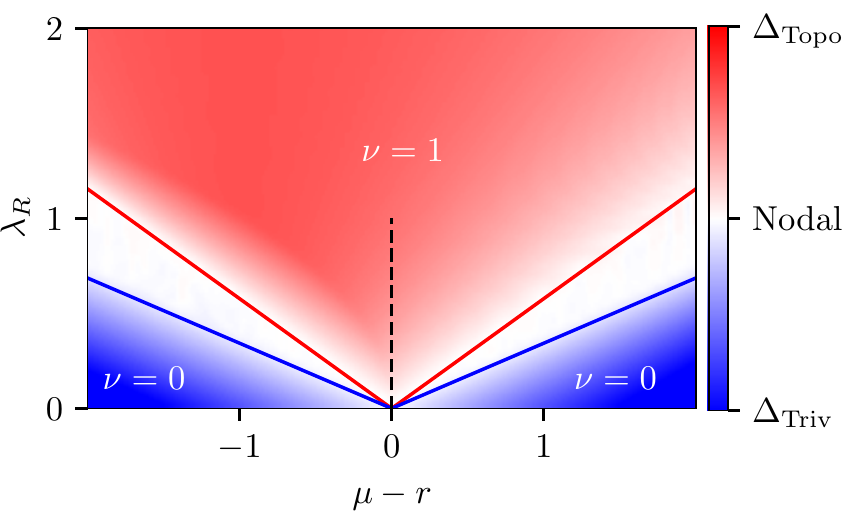}
\caption{Phase diagram of a {\it clean} hybrid structure consisting of an Fe-based $s_\pm$-wave superconductor and a 2D Rashba SOC layer, described by Eq.\,(\ref{tight-binding.eq}), as a function of $\mu-r$ and $\lambda_R$. Topologically trivial $s_{++}$-wave ($\nu=0$), non-trivial $s_{\pm}$-wave ($\nu=1$), and nodal phases are depicted by blue, red, and white, respectively. The color intensity coding corresponds to the energy gap size. Blue and red solid lines mark the phase boundaries.
}
\label{PD0.fig}
\end{figure}

In Fig.\,\ref{PD0.fig} we present the phase diagram of the Hamiltonian Eq.\,(\ref{tight-binding.eq}) as a function of the Rashba SOC $\lambda_R$ and chemical potential $\mu$. Depending on the size of $\lambda_R$ and $\mu$ compared to the ratio of the superconducting gaps, $r=t\Delta_0/\Delta_1$, the system has three different phases: topologically trivial ($\nu = 0$), non-trivial ($\nu = 1$), and nodal phase \cite{Zhang13}. The topological phase occurs for ${|\mu-r|<2\lambda_R\sqrt{|r|-r^2/4}}$ (red in Fig.~\ref{PD0.fig}), while the trivial phase requires ${|\mu-r|>2\lambda_R\sqrt{2-r^2/8}}$ (blue in Fig.~\ref{PD0.fig}). The white region in between these two phases represents the nodal phase, where at least one of the SOC split Fermi surfaces crosses the nodal line of the superconducting gap function.
Interestingly, the trivial, topological, and nodal phases all meet at the origin $\mu=r,\lambda_R=0$, forming a triple point in the phase diagram.

To include the effect of non-magnetic, or potential, disorder, we independently implement two different disorder models, namely local random chemical potential fluctuations, or Anderson disorder, and dilute but strong non-magnetic scatterers, or concentration disorder. Anderson disorder is modeled by adding an on-site random chemical potential $\epsilon_{\bf{i}}$ to every site,
\begin{equation}
{\cal H}^{(1)}_{dis}
 = \sum_{ {\bf{i}} \sigma} \epsilon_{\bf{ i}}  c_{{\bf i}\sigma}^\dag c_{{\bf i}\sigma}
\label{dis_Ham.eq}
\end{equation}
where $\epsilon_{\bf{i}}$ is chosen from a box distribution, $\epsilon_{\bf{i}} \in [-W/2,W/2]$. To simulate concentration disorder, we add a constant strong potential to a randomly chosen, small, subset of lattice sites,
\begin{equation}
{\cal H}^{(2)}_{dis}
 = \sum_\sigma \sum_{ {\bf{i}} \in \Lambda^\ast} V  c_{{\bf i}\sigma}^\dag c_{{\bf i}\sigma}
\label{dis_Ham.eq2}
\end{equation}
where $\Lambda^\ast$ is a small subset of all sites of the square lattice $\Lambda$. The disorder concentration $c$ is then given by the ratio of the dimensions of $\Lambda^\ast$ and $\Lambda$, which becomes a chosen input parameter, while the position of the sites contained in $\Lambda^\ast$ is random. A notable difference between these two disorder types is that, for Anderson disorder the added on-site terms average to zero, while for concentration disorder, the effective chemical potential is shifted, on average, by $cV$.

We solve ${ {\cal H} = {\cal H}_0 + {\cal H}^{(i)}_{dis}}$ using exact diagonalization within the TBTK code \cite{Bjornson2013,Bjornson2015,Bjornson2016,KristoferTBTK}. We use large lattices with both periodic boundary conditions  (PBC) and open boundary conditions (OBC) in order to both assess the effect of disorder on bulk properties and its influence on the topological edge states appearing at the boundary in the topological phase. To be able to compare our results for Eq.~\eqref{tight-binding.eq} with those of a conventional $s$-wave superconductor, we also perform the same calculations but setting $\Delta_1=0$. This results in modeling a  trivial and conventional $s$-wave superconductor with only onsite pairing, which has no topological features and benefits from strong protection against disorder as established by Anderson's theorem \cite{Anderson59}.

In this work, for simplicity, we assume $\Delta_0=\Delta_1=0.2$ and $\lambda_R = 0.5$. We also choose $\mu = 1.0$ for representing the topological $s_\pm$-wave phase and  $\mu = 2.9$ for representing the trivial $s_{++}$-wave phase, unless otherwise stated. For these two values of the chemical potential, the size of the excitation gap in the clean case is almost the same, $2\Delta \approx 0.24$, and we can thus directly compare the influence of disorder in the subgap regime between these two topologically distinct cases. For the conventional $s$-wave superconductor we set $\Delta_0=0.15$, which gives rise to an excitation gap similar to the two aforementioned superconducting states. Note that we here keep the superconducting order parameter constant throughout the sample. Technically, a more accurate treatment would require us to calculate $\Delta_0$ and $\Delta_1$ in a self-consistent manner. However, previous tests on similar proximity-induced hybrid structures have revealed only minor quantitative corrections to the gap size and hardly any changes to phase diagram\,\cite{BlackSchaffer11, Mascot19}.

Finally, finite size effects are usually unavoidable in solid state simulations. The inclusion of random disorder makes it even more challenging to obtain configuration-independent results. Throughout this work we report exact diagonalization results using a square lattice with $51 \times 51$ sites  (unless otherwise stated), but we have checked our results also using smaller lattice sizes.
Moreover, when we perform disorder averaging we choose the number of disorder realizations $n$ such that ${\sigma}/{\sqrt{n}} < 0.015$ is fulfilled, where $\sigma$ is the standard deviation. 
Typical values here are $n=40$ and $\sigma = 0.02$. This procedure guarantees that all our disorder averaged results are sufficiently independent of a specific disorder realization. Furthermore, in order to calculate the bulk DOS we use a Gaussian kernel for smoothing the data, setting the standard deviation for the Gaussian kernel to $\Gamma=0.02$.

%
%

\section{Results}
\label{Results}
Having established the phase diagram and the overall properties of our system in the clean limit we now turn to the effects of non-magnetic disorder in this TRI topological superconductor. First, we study the properties of the bulk DOS in the presence of different disorder types and strengths. Despite naive expectations of disorder robustness for topological systems as long as their symmetries are preserved, we show that the TRI topological superconducting phase is actually very fragile against non-magnetic disorder. This leads us to derive a very different phase diagram at finite disorder, where especially the nodal phase is heavily expanded.
Finally, we explore the impact of disorder on the helical Majorana edge states where we capture their delocalization and eventual destruction with increasing disorder.  

\subsection{Bulk density of states}
\label{DOS}
\begin{figure}[t!]
\centering
\includegraphics[width=\columnwidth]{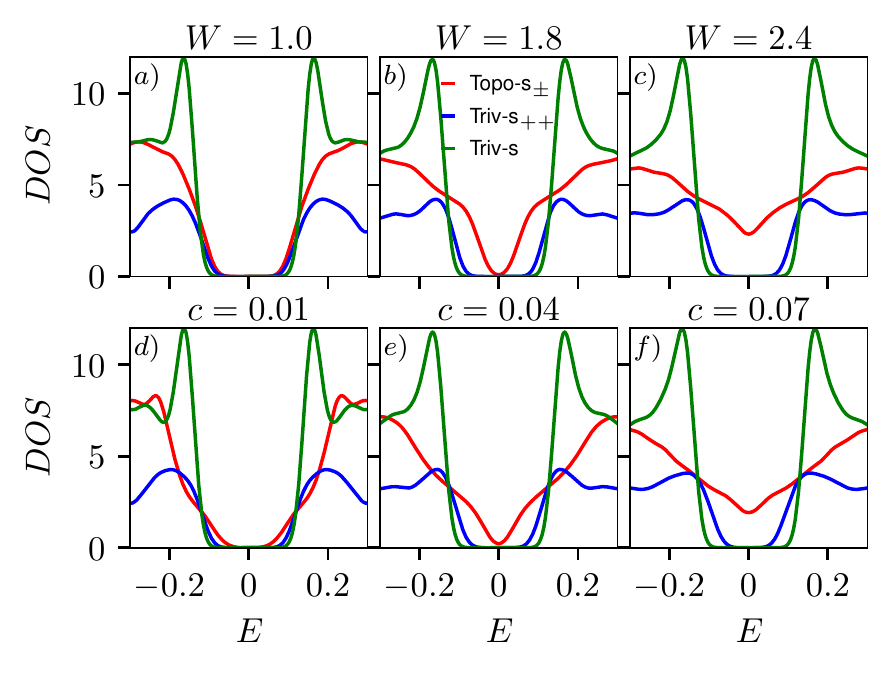}
\caption{Bulk DOS for different superconducting phases in the presence of disorder with topological $s_\pm$-wave (red), trivial $s_{++}$-wave (blue), and conventional $s$-wave (green) superconducting symmetries, all with similar energy gaps in the clean limit. Top row: Anderson disorder with $W=1.0, 1.8$, and $2.4$. Bottom row:  concentration disorder with $V=2$ for $c=1\%, 4\%$, and $7\%$. 
}
\label{3+3.fig}
\end{figure}

To illustrate the impact of non-magnetic disorder on the bulk properties of a TRI topological superconductor, we first evaluate the bulk DOS for the Hamiltonian Eq.~(\ref{tight-binding.eq}) in the presence of both Anderson and concentration disorder, $\mathcal{H}_{dis}^{(i)}$ ($i=1,2$), with PBCs imposed. In Fig.~\ref{3+3.fig} we show the DOS for a single disorder configuration for three different disorder strengths $W$ (top) and disorder concentrations $c$ (bottom).
Each panel in Fig.\,\ref{3+3.fig} shows the DOS for three different superconducting states with topological $s_\pm$-wave (red), trivial $s_{++}$-wave (blue), and trivial conventional onsite $s$-wave (green) symmetries.  
In particular, in Figs.\,\ref{3+3.fig}\,(a-c), we consider Anderson disorder using $W= 1.0, 1.8$ and $2.4$, respectively. Initially, in the absence of the disorder and for only small disorder, the size of the energy gap is equal for all three superconducting states, see Fig.\,\ref{3+3.fig}\,(a). However, when increasing the strength of the Anderson disorder, we find the DOS to be different between the topological and trivial phases in a number of respects. Most importantly, the gap size in the topological $s_\pm$-wave superconducting phase is strongly reduced. Surprisingly, even for $W\approx1.8$, which is still small in comparison to the superconducting band width of the  Hamiltonian $D \approx 15$, the gap of the topological phase closes, see Fig.\,\ref{3+3.fig}\,(b). Increasing the disorder further, what was the topological phase becomes gapless, see Fig.\,\ref{3+3.fig}\,(c). In contrast, the conventional $s$-wave state is robust for all reported disorder strengths without any notable change in the gap size. Similarly, for the trivial $s_{++}$-wave state the energy gap stays wide open only showing a minimal reduction in the presence of disorder, at least as long as $W \leq D$.

Next, we investigate concentration disorder in Fig.\,\ref{3+3.fig}\,(d-f). Here, we set  the impurity scattering strength to $V=2$ and vary the disorder concentration between $1\%, 4\%,$ and $7\%$, respectively. Again, we observe that very small concentrations of disorder strongly affect the topological phase. As seen in Fig.~\ref{3+3.fig}(e),  an impurity concentration of only $4\%$ fully closes the energy gap of the topological phase, producing a gapless state. In contrast, the conventional $s$-wave state remains completely unaffected even for large concentrations. Also the trivial $s_{++}$-wave state survives large impurity concentrations, albeit with a slightly reduced gap size. To conclude, the results in Fig.~\ref{3+3.fig} show that for both Anderson and concentration disorder the topological phase is very fragile against disorder, in sharp contrast to the robustness of trivial $s$-wave phases. 
Given that the disorder does not break any of the symmetries protecting the topological phase, this is an unexpected finding.

\begin{figure}[tb]
\center
\includegraphics[width=\columnwidth]{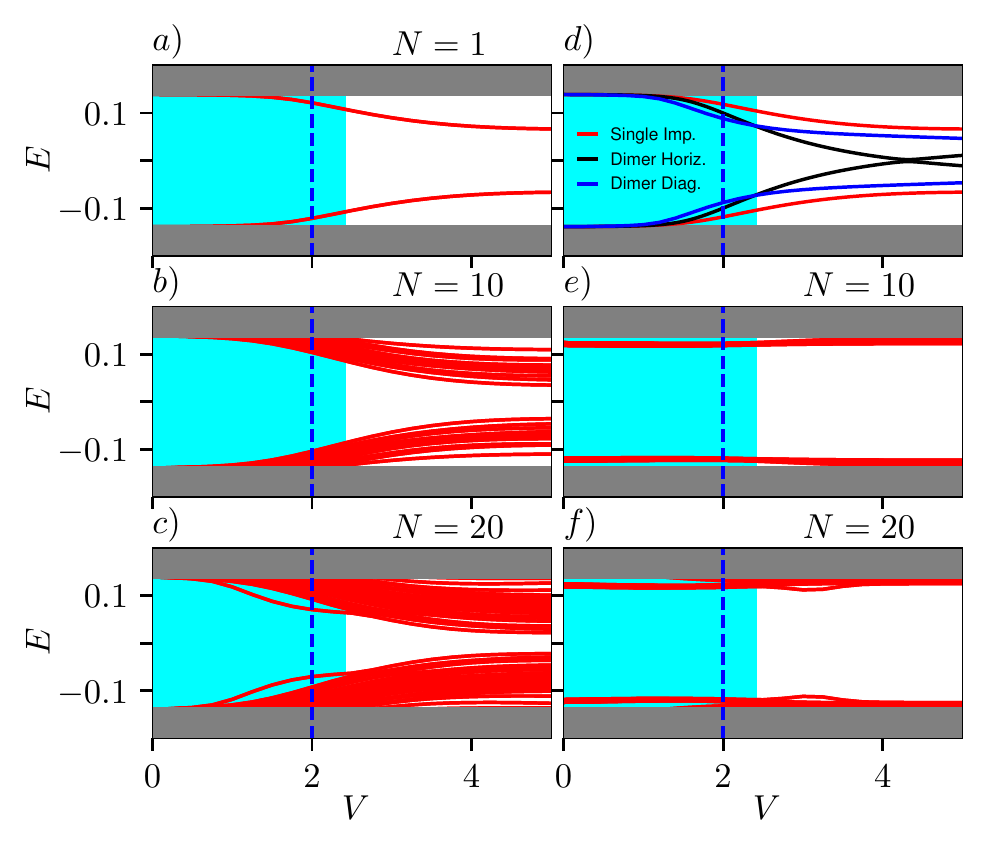}
\caption{Energy of subgap states induced by $N$ randomly placed non-magnetic impurities on a $51\times 51$ lattice as a function of impurity scattering strength $V$. Grey shade corresponds to the bulk bands in the clean limit.
Cyan shade and dashed blue line mark the disorder strengths for the Anderson and concentration disorder used in Fig.~\ref{3+3.fig}.
(a-d) Topological $s_\pm$-wave phase for (a) $N=1$, (b) $N=10$, (c) $N=20$ (d) single impurity vs.\ dimers of impurities}. 
(e-f) Trivial $s_{++}$-wave phase for (e) $N=10$ and (f) $N=20$. 
(d) Single impurity and single impurity dimer oriented either horizontally or diagonally.
\label{subgap.fig}
\end{figure}

We can start to understand the strong effect of non-magnetic disorder on the topological phase and its gap by first considering the single-impurity case.
As shown in Ref.~\cite{Mashkoori2018}, a single potential impurity added to the TRI superconductor in Eq.~\eqref{tight-binding.eq} produces a single pair of subgap states (symmetric around zero energy and two-fold degenerate due to time-reversal symmetry) in the energy spectrum in the topological phase, while the trivial phase never hosts any subgap states for non-magnetic impurities. However, a single impurity was also shown to never induce states at or near zero energy. In our results, we clearly see that we generate a finite bulk DOS also at zero energy for a finite concentration of disorder. 

To connect the earlier single impurity results \cite{Mashkoori2018} to our bulk results, we next analyze the energy spectrum when we add $N$ randomly positioned impurities, letting $N=1, \ldots, 20$. The results are summarized in Fig.\,\ref{subgap.fig} where we plot the subgap spectrum as a function of the impurity strength $V$. 
The case of a single impurity $N=1$ is shown in Fig.\,\ref{subgap.fig}\,(a), where we observe subgap state clearly forming for $V\gtrsim 1.5$ but notably always staying rather far from zero energy. With increasing number of impurities, $N$, we observe an increasing number of subgap states that also move closer to zero energy. In Fig.\,\ref{subgap.fig}\,(b,c) we plot the cases of $N=10$ and $N=20$, respectively. Already for $N=20$, which corresponds to an impurity concentration of only $c=0.008$, we observe a considerable contraction of the energy gap. 
From the plotted sequence of disorder sites, $N=1, 10, 20$, we can easily extrapolate to the results for larger concentrations obtained in Figs.~\ref{3+3.fig}(a-f) using $V=2$, marked by a vertical dashed blue line in Fig.~\ref{subgap.fig}. In this way we obtain a smooth connection between the single-impurity results and bulk results at finite disorder concentration, noting that the difference is that multiple impurities pushes the subgap impurity states closer and closer to zero energy, eventually filling the whole energy gap.

To contrast the results for the topological phase, we also plot results for $N=10, 20$ impurities in the trivial $s_{++}$-wave phase in Fig.\,\ref{subgap.fig}\,(e,f). For a few impurities, we hardly observe any subgap states, only for $N\sim 10$ do subgap states start to slowly leak into the bulk gap, but they always remain extremely close to the bulk band edge. This explains the slightly reduced bulk gap in Figs.~\ref{3+3.fig}(a-f) in the trivial phase. Finally, in Fig.\,\ref{subgap.fig}\,(d) we consider the special case of an ``impurity dimer'', i.e., two impurities next to each other and of identical strength. The dimer generally produces lower-lying impurity subgap states than the single impurity, but the exact values vary depending on the dimer orientation on the lattice. In fact, horizontally bonded dimers even achieve zero-energy states at certain disorder strengths. This further supports the fact that for random finite disorder concentration, zero and near zero energy states will always be present.

If we instead redo the same calculation as in Fig.~\ref{subgap.fig} but use a disorder strength that on each disorder site varies between $[-V,V]$, we arrive at qualitatively the same results, which eliminates any possible influence of the shift of the overall chemical potential.
We also note that all plots shown in Fig.\,\ref{subgap.fig} represent only a single disorder realization, but we do not expect disorder-averaging over multiple realizations to qualitatively change the results.

\begin{figure}[t]
\centering
\includegraphics[width=\columnwidth]{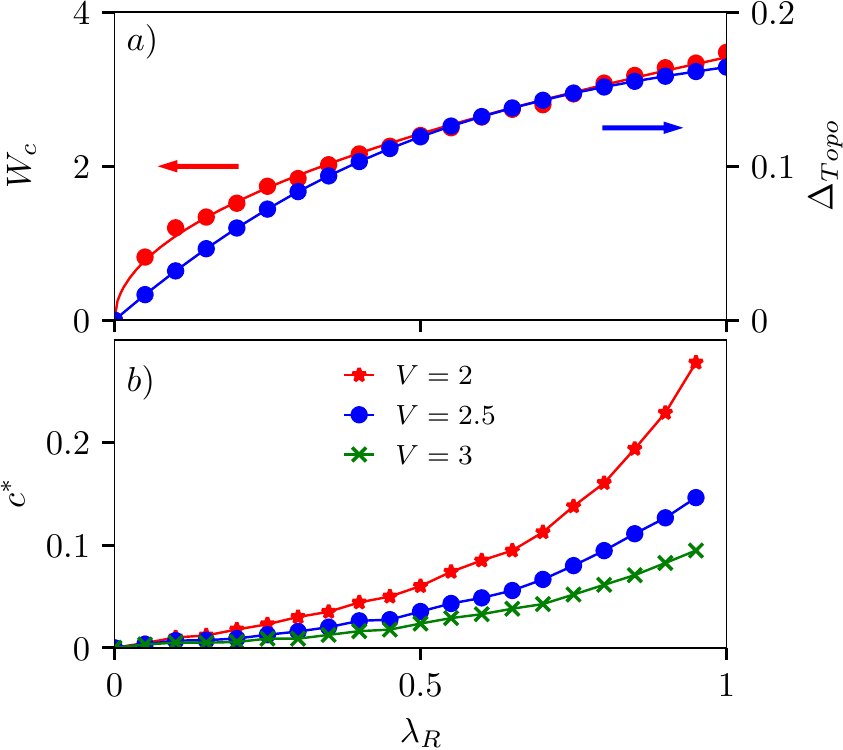}
\caption{(a) Critical disorder strength $W_c$ of Anderson disorder (red, left axis) and topological gap $\Delta_{\rm Topo}$ in the clean limit (blue, right axis) as a function of $\lambda_R$. (b) Critical concentration $c^*$ for concentration disorder as a function of $\lambda_R$.  Solid red line in (a) is a fit to $W_c\propto \lambda_R^{1/2}$, all other lines are only guides to the eye.}
\label{WcAd.fig}
\end{figure}

To shed further light on the effects of disorder we investigate more closely when the topological gap is lost. To this end, we introduce $W_c$, the ensemble-averaged disorder strength for which the transition from gapped topological to gapless phase takes place for Anderson disorder. To be computationally achievable, we choose to evaluate $W_c$ for points in the topological phase along the dashed line in the phase diagram in Fig.\,\ref{PD0.fig} and for $\lambda_R \in [0,1]$, which is the most physically relevant range. In the clean system, the phase diagram is geometrically symmetric along $\mu=r$ and the gap size in the topological phase, $\Delta_\mathrm{Topo}$, grows with increasing $\lambda_R$ as shown by the blue curve in Fig.\,\ref{WcAd.fig}\,(a). By plotting $W_c$ as a function of $\lambda_R$ in red and in the same panel, we find that $W_c$ has almost exactly the same dependence as the gap $\Delta_\mathrm{Topo}$ on $\lambda_R$. In more detail, we observe an algebraic relation, $W_c\propto \lambda_R^{1/2}$, as indicated by the fit (solid red line) in Fig.\,\ref{WcAd.fig}(a).
This strongly suggests that $\Delta_\mathrm{Topo}$ is the relevant energy scale for the transition to the gapless phase. Hence, the larger the topological gap in the clean limit, the more robust is the topological phase, as also intuitively expected. This means that in regions close to the nodal phase in the clean limit in Fig.~\ref{PD0.fig}, even very weak disorder will destroy the topological gap and render a nodal spectrum.

We can alternatively evaluate the critical density $c^*$, for which concentration disorder completely fills the topological gap and generates a nodal phase. In Fig.~\ref{WcAd.fig}(b) we plot $c^*$ as a function of $\lambda_R$ using several different values of $V$. We find again that stronger disorder, here represented by higher disorder densities, are needed to reach the critical density for larger $\lambda_R$, or, equivalently, a larger topological gap $\Delta_\mathrm{Topo}$ in the clean limit.
Instead keeping $\lambda_R$ constant, for larger $V$, we find smaller $c^*$, since stronger scatterers pushes the impurity subgap states further into the gap as seen in Fig.~\ref{subgap.fig}, which facilities the transition to the nodal phase.

\begin{figure}[t]
\centering
\includegraphics[width=\columnwidth]{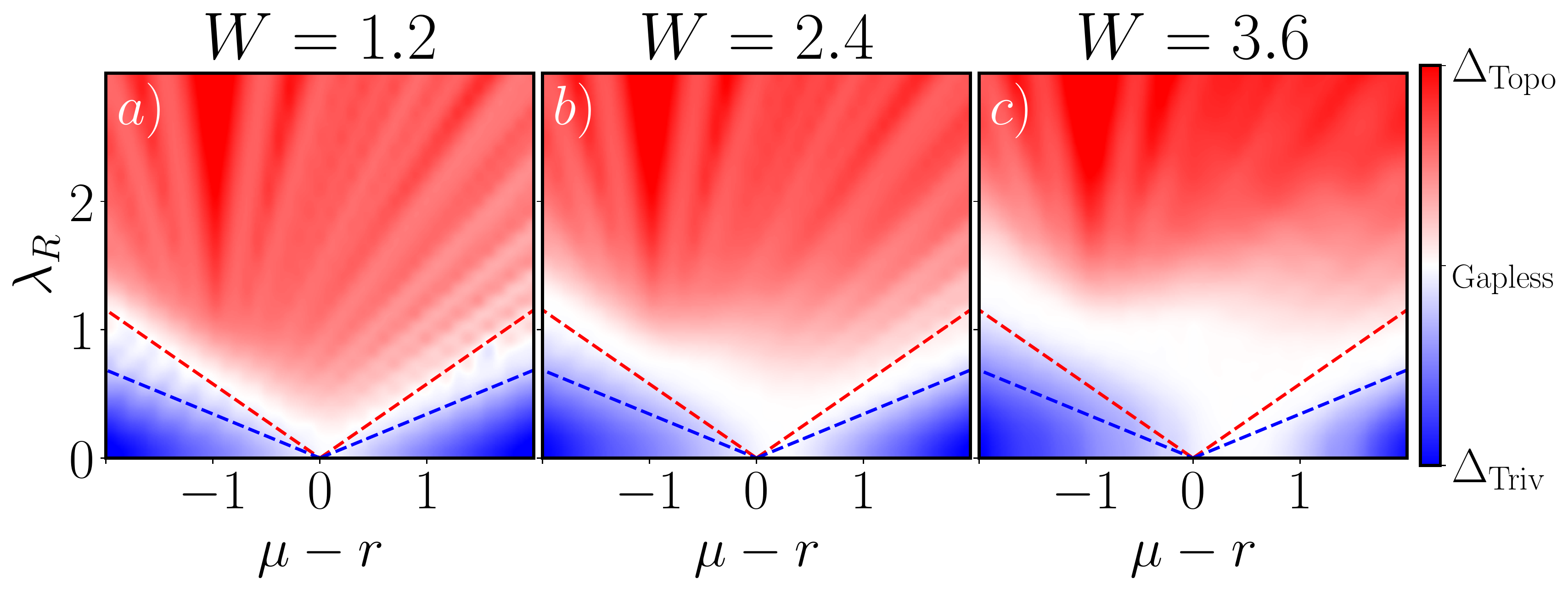}
\caption{Phase diagram as a function of Rashba SOC $\lambda_R$ and chemical potential $\mu-r$ for Anderson disorder with $W=1.2$ (a), $W=2.4$ (b), and $W=3.6$ (c), representing disordered versions of the clean phase diagram Fig.\,\ref{PD0.fig}.
Blue and red dashed lines separate different phases in the clean system.}
\label{PD.fig}
\end{figure}

\subsection{Phase diagram}
\label{PhaseDiagram}
Having established that the energy gap in the topological phase is quickly diminished and even disappears in the presence of disorder, we next establish the phase diagrams for finite disorder.
By using extensive numerical simulations extracting both the topology and gap size, we produce in Fig.\,\ref{PD.fig}(a-c) the phase diagrams analogous to the clean system in Fig.\,\ref{PD0.fig} for different strengths  $W$ of Anderson disorder, averaged over $50$ independent disorder configurations. As expected, stronger disorder affects the phase diagram drastically. First of all, the nodal region (white color) grows substantially and especially occupies regions that were topological in the clean limit. Second, the trivial phase is much less affected by disorder than the topological phase. Additionally, while the nodal phase shrinks in the clean limit into a triple point at $\mu=r$ and $\lambda_R=0$, in the presence of disorder we find the nodal phase expanding over a large area for small $\lambda_R$.
We also note that disorder affects the phase diagram in an asymmetric way with respect to $\mu-r$. This is not surprising as the gap size in the clean limit is not symmetric with respect to $\mu-r$, see color intensity in Fig.\,\ref{PD0.fig}. As a consequence, the filling of the gap through disorder-induced states occurs differently for positive and negative values of $\mu-r$.

\begin{figure}[t]
\centering
\includegraphics[width=1.0\columnwidth]{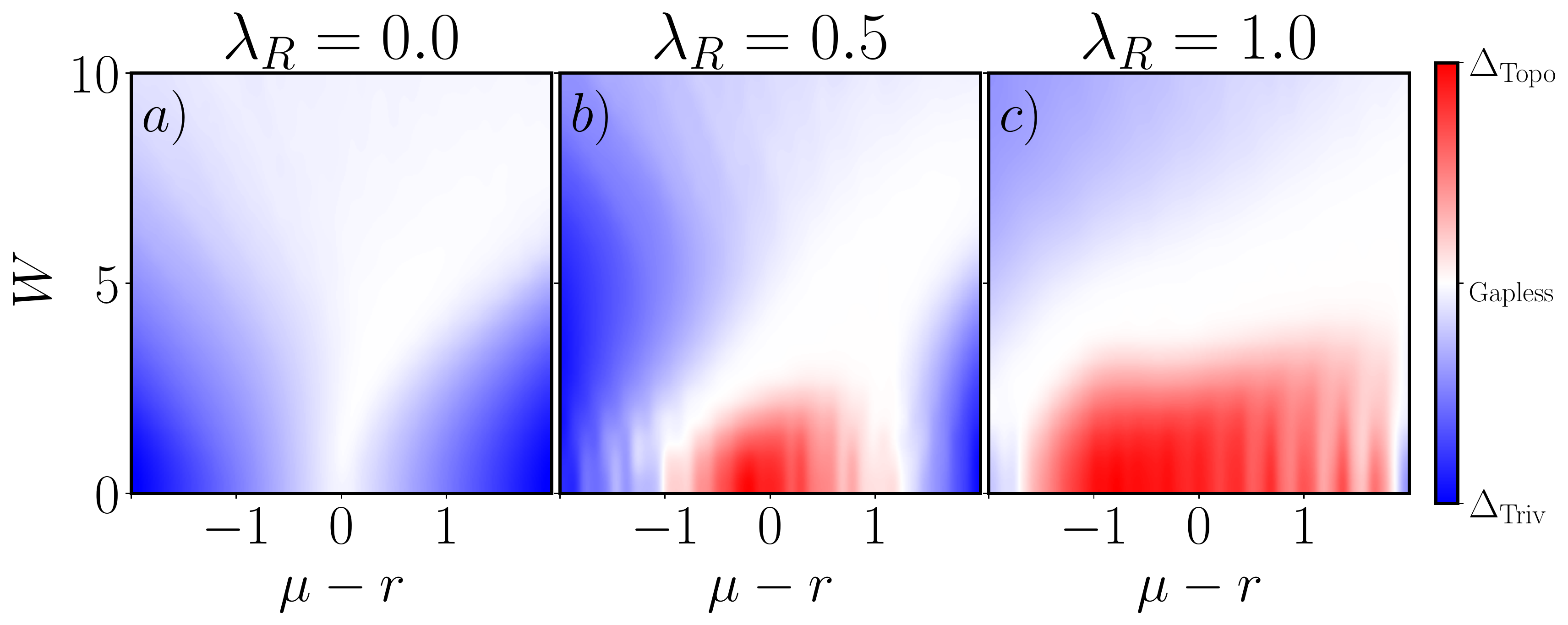}
\caption{Phase diagram as a function of disorder strength $W$ and chemical potential $\mu-r$ for $\lambda_R=0$ (a), $\lambda_R=0.5$ (b), and $\lambda_R=1.0$ (c).}
\label{PDW.fig}
\end{figure}

To further illustrate the effect of disorder on the phase diagram, we plot in Fig.~\ref{PDW.fig} phase diagrams for fixed Rashba SOC $\lambda_R$, while we tune the chemical potential and disorder strength. As shown in Fig.~\ref{PDW.fig}(a), there is no topological phase for $\lambda_R=0$. For moderate disorder strength, the trivial phase also remains gapped. By further increasing the disorder strength, the gap is however reduced and eventually, for disorder of the order of the bandgap, we lose the gap even in the trivial phase. 
Next setting $\lambda_R = 0.5$ in Fig.~\ref{PDW.fig}(b), the clean system is topological for ${|\mu-r|\lessapprox 1}$. By adding disorder, we clearly see how the topological phase is much more fragile to disorder than the trivial phase. Finally, in Fig.~\ref{PDW.fig}(c), we set $\lambda_R=1.0$ which in the clean limit generates a topological phase for the whole range of ${|\mu-r| \lessapprox 2}$. Again, we observe that disorder leads to strong suppression of the gap in the topological phase, but due to the initially larger topological gap thanks to the larger $\lambda_R$, the transition to the nodal phase requires a larger $W_c$ compared to Fig.~\ref{PDW.fig}(b).
Somewhat surprisingly, we find that  further increasing the disorder strength $W$ beyond $W_c$ in Fig.\,\ref{PDW.fig}\,(b-c) eventually drives the system from the gapless nodal phase into the trivially gapped phase for a range of $\mu -r$. By using OBC and calculating the local DOS we have  verified that there are no edge states in this regime, consistent with the trivial topology. 
This behavior of a re-entrant gapped phase is reminiscent of  disorder-induced topological phase transitions found earlier in superconductors, where Anderson disorder has been shown to lead to topological phase transitions between phases with different Chern numbers \cite{Borchmann16}.

\begin{figure}[t!]
\centering
\includegraphics[width=1.\columnwidth]{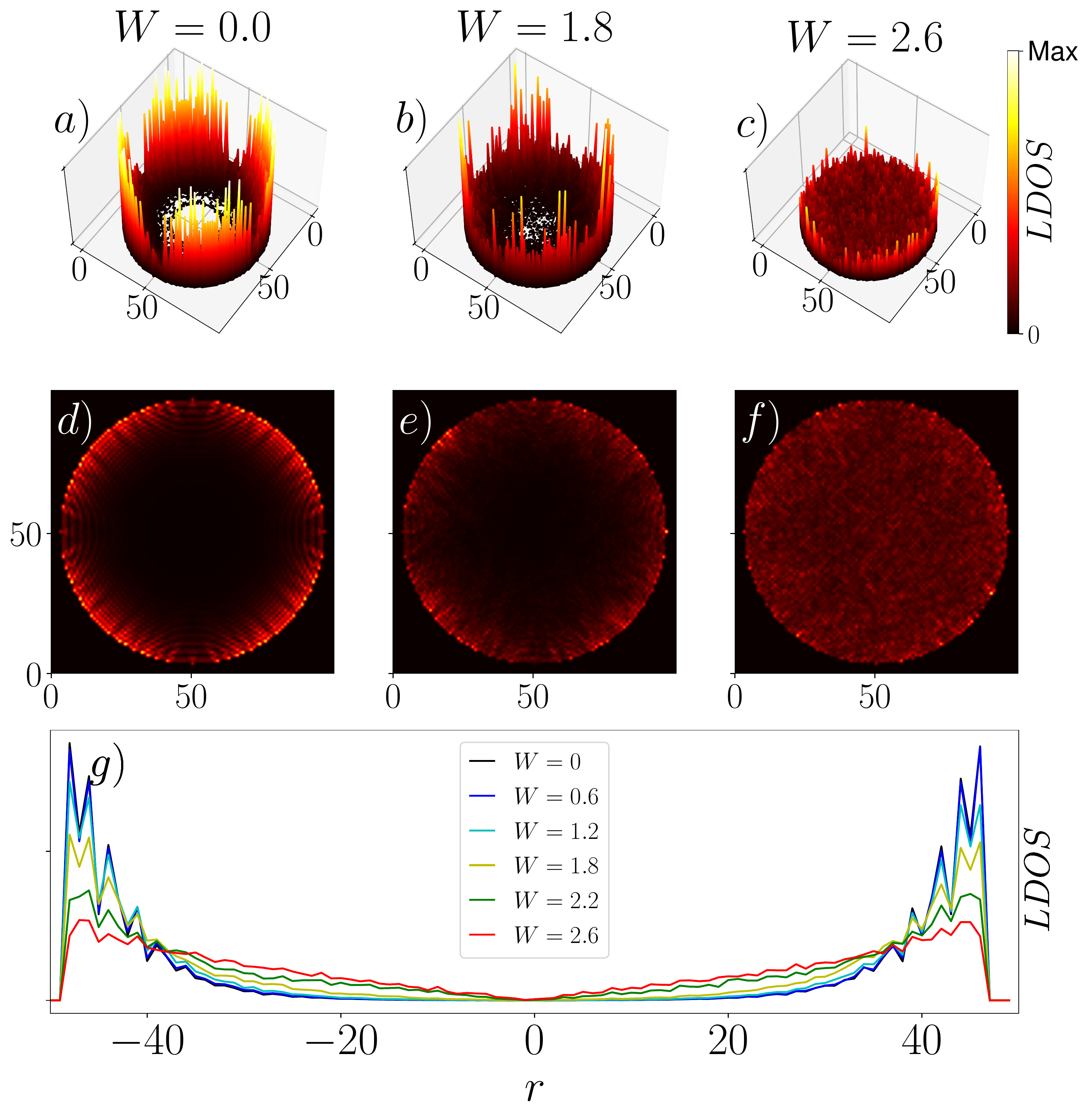}
\caption{Local DOS for subgap energies $|E|< \Delta_\mathrm{Topo}/4$ in the topological phase for a finite disc of radius $r=47$. 2D and 3D plots of the local DOS in the clean limit (a,d) and for $W=1.8$ (b,e) and $W=2.6$ (c,f).
(g) Line cuts of the local DOS in same low-energy range through the disc for different disorder strengths $W$.}
\label{LDOS_tuneDelta.fig}
\end{figure}

\subsection{Edge states}
\label{EdgeStates}
To further investigate the detrimental effect of disorder on topological $s_\pm$-wave TRI superconductors, we study the disorder impact on the helical Majorana edge modes by performing OBC calculations for a disc geometry with radius $r=47$ lattice points. Such a disc size is sufficient to avoid overlapping of the edge wave functions at opposite sides of the disc, which would lead to their hybridization. 
By tuning the Anderson disorder strength $W$, we observe a clear destructive influence of disorder on the edge modes. As an illustration, we plot in Fig.~\ref{LDOS_tuneDelta.fig} the accumulated local DOS for subgap energies $|E|<\Delta_\mathrm{Topo}/4$ using a specific disorder configuration. Fig.\,\ref{LDOS_tuneDelta.fig}\,(a-c) and (d-f) contain identical data but plotted in 3D and 2D, respectively. In Fig.\,\ref{LDOS_tuneDelta.fig}\,(a,d), in the absence of disorder, we  find well-localized helical Majorana edge modes and a fully gapped interior. By introducing moderately small Anderson disorder, as shown in Fig.\,\ref{LDOS_tuneDelta.fig}\,(b,e), the helical Majorana edge modes already begin to fade away and near-zero subgap states begin to be visible throughout the disc interior. When the disorder strength is increased further, see Fig.\,\ref{LDOS_tuneDelta.fig}\,(c,f), the edge modes are strongly suppressed and a clear subgap DOS develops in the interior. 
Similar calculations for concentration disorder confirm that the helical Majorana edge modes in the topological phase disappear with increasing impurity concentration.
As a complement, we also plot line cuts through the disc in Fig.\,\ref{LDOS_tuneDelta.fig}\,(g). Here it becomes clear how the delocalization of the edge states goes hand in hand with the emergence of subgap states in the bulk. This result shows how the loss of the topological gap in the bulk directly affects the edge states, which fade into the bulk and eventually disappear with increasing disorder.
In other words, the edge of the topological phase is affected by  disorder in the same way as the bulk, a manifestation of the bulk-boundary correspondence.

%
%

\section{Summary}
\label{summary}
In this work, we study the effects of non-magnetic disorder in a time-reversal invariant superconductor, which can be realized in a hybrid structure consisting of an Fe-based superconductor and a spin-orbit coupled Rashba layer. The hybrid structure hosts three different phases, a topological $s_\pm$-wave and a trivial $s_{++}$-wave, with an intersecting nodal phase. We perform intensive numerical calculations in the presence of two different types of disorder, Anderson disorder, i.e., random chemical potential fluctuations, and concentration disorder, i.e., random dilute strong potential scatterers, both preserving time-reversal symmetry. The naive expectation is that the topological phase should be robust against this disorder as it preserves all symmetries protecting the topology.
Instead we find that the disorder leads to subgap states quickly accumulating in the gap of the topological phase, leading to the closure of the topological gap even for moderately weak disorder. 
These findings are in contrast to the topologically trivial phase, as well as conventional $s$-wave superconductors, which are both exceptionally robust to disorder with no notable subgap states even for strong disorder. We are able to trace this disorder fragility of the topological phase to the existence of subgap states in the single- and few-impurity limit. Notably though, single impurities do not generate (near) zero-energy states, while random disorder fully closes the gap, generating a gapless phase.
We derive the phase diagram in the presence of disorder, where we very generally observe the expansion of the nodal phase at the expense of the topological phase.
In addition to a bulk analysis, we also investigate the helical Majorana edge modes associated with the topological phase. We find that even moderately weak disorder causes the edge states to decay into the bulk and eventually to disappear. Thus, we find that disorder affects both edge and bulk of the topological phase in the same detrimental way. 
Our results suggest that the lack of a hard energy gap and the extension of the nodal phase due to disorder can easily prevent experimental detection of time-reversal invariant topological superconducting phases.

%
%
\acknowledgments
We thank M.~Shiranzaee for fruitful discussions and acknowledge financial support from the Swedish Research Council (Vetenskapsradet Grant No. 2018-03488), the Knut and Alice Wallenberg Foundation through the Wallenberg Academy Fellows program, and the Australian Research Council through Grant No.\ DP200101118. Calculations were enabled by resources provided by the Swedish National Infrastructure for Computing (SNIC) at UPPMAX, partially funded by the Swedish Research Council through Grant No.~2018-05973. 
 
%

%
%

\bibliography{bibFile_2.bib}

\begin{thebibliography}{44}%
\makeatletter
\providecommand \@ifxundefined [1]{%
 \@ifx{#1\undefined}
}%
\providecommand \@ifnum [1]{%
 \ifnum #1\expandafter \@firstoftwo
 \else \expandafter \@secondoftwo
 \fi
}%
\providecommand \@ifx [1]{%
 \ifx #1\expandafter \@firstoftwo
 \else \expandafter \@secondoftwo
 \fi
}%
\providecommand \natexlab [1]{#1}%
\providecommand \enquote  [1]{``#1''}%
\providecommand \bibnamefont  [1]{#1}%
\providecommand \bibfnamefont [1]{#1}%
\providecommand \citenamefont [1]{#1}%
\providecommand \href@noop [0]{\@secondoftwo}%
\providecommand \href [0]{\begingroup \@sanitize@url \@href}%
\providecommand \@href[1]{\@@startlink{#1}\@@href}%
\providecommand \@@href[1]{\endgroup#1\@@endlink}%
\providecommand \@sanitize@url [0]{\catcode `\\12\catcode `\$12\catcode
  `\&12\catcode `\#12\catcode `\^12\catcode `\_12\catcode `\%12\relax}%
\providecommand \@@startlink[1]{}%
\providecommand \@@endlink[0]{}%
\providecommand \url  [0]{\begingroup\@sanitize@url \@url }%
\providecommand \@url [1]{\endgroup\@href {#1}{\urlprefix }}%
\providecommand \urlprefix  [0]{URL }%
\providecommand \Eprint [0]{\href }%
\providecommand \doibase [0]{http://dx.doi.org/}%
\providecommand \selectlanguage [0]{\@gobble}%
\providecommand \bibinfo  [0]{\@secondoftwo}%
\providecommand \bibfield  [0]{\@secondoftwo}%
\providecommand \translation [1]{[#1]}%
\providecommand \BibitemOpen [0]{}%
\providecommand \bibitemStop [0]{}%
\providecommand \bibitemNoStop [0]{.\EOS\space}%
\providecommand \EOS [0]{\spacefactor3000\relax}%
\providecommand \BibitemShut  [1]{\csname bibitem#1\endcsname}%
\let\auto@bib@innerbib\@empty
\bibitem [{\citenamefont {Wen}(2004)}]{WenBook}%
  \BibitemOpen
  \bibfield  {author} {\bibinfo {author} {\bibfnamefont {X.}~\bibnamefont
  {Wen}},\ }\href {https://books.google.se/books?id=RYESDAAAQBAJ} {\emph
  {\bibinfo {title} {Quantum Field Theory of Many-Body Systems: From the Origin
  of Sound to an Origin of Light and Electrons}}},\ Oxford Graduate Texts\
  (\bibinfo  {publisher} {OUP Oxford},\ \bibinfo {year} {2004})\BibitemShut
  {NoStop}%
\bibitem [{\citenamefont {Bernevig}\ and\ \citenamefont
  {Hughes}(2013)}]{BernevigBook}%
  \BibitemOpen
  \bibfield  {author} {\bibinfo {author} {\bibfnamefont {B.}~\bibnamefont
  {Bernevig}}\ and\ \bibinfo {author} {\bibfnamefont {T.}~\bibnamefont
  {Hughes}},\ }\href {https://books.google.se/books?id=\_7r\_UqFN0IEC} {\emph
  {\bibinfo {title} {Topological Insulators and Topological Superconductors}}}\
  (\bibinfo  {publisher} {Princeton University Press},\ \bibinfo {year}
  {2013})\BibitemShut {NoStop}%
\bibitem [{\citenamefont {Moessner}\ and\ \citenamefont
  {Moore}(2021)}]{Moessner2021}%
  \BibitemOpen
  \bibfield  {author} {\bibinfo {author} {\bibfnamefont {R.}~\bibnamefont
  {Moessner}}\ and\ \bibinfo {author} {\bibfnamefont {J.~E.}\ \bibnamefont
  {Moore}},\ }\href {https://books.google.de/books?id=BTMiEAAAQBAJ} {\emph
  {\bibinfo {title} {{Topological Phases of Matter}}}}\ (\bibinfo  {publisher}
  {Cambridge University Press},\ \bibinfo {year} {2021})\BibitemShut {NoStop}%
\bibitem [{\citenamefont {Kitaev}(2001)}]{Kitaev01}%
  \BibitemOpen
  \bibfield  {author} {\bibinfo {author} {\bibfnamefont {A.~Y.}\ \bibnamefont
  {Kitaev}},\ }\href {\doibase 10.1070/1063-7869/44/10S/S29} {\bibfield
  {journal} {\bibinfo  {journal} {Phys. Usp.}\ }\textbf {\bibinfo {volume}
  {44}},\ \bibinfo {pages} {131} (\bibinfo {year} {2001})}\BibitemShut
  {NoStop}%
\bibitem [{\citenamefont {Wilczek}(2009)}]{Wilczek09}%
  \BibitemOpen
  \bibfield  {author} {\bibinfo {author} {\bibfnamefont {F.}~\bibnamefont
  {Wilczek}},\ }\href {\doibase 10.0.4.14/nphys1380} {\bibfield  {journal}
  {\bibinfo  {journal} {Nat. Phys.}\ }\textbf {\bibinfo {volume} {5}},\
  \bibinfo {pages} {614} (\bibinfo {year} {2009})}\BibitemShut {NoStop}%
\bibitem [{\citenamefont {Aguado}(2017)}]{Aguado2017}%
  \BibitemOpen
  \bibfield  {author} {\bibinfo {author} {\bibfnamefont {R.}~\bibnamefont
  {Aguado}},\ }\href {\doibase 10.1393/ncr/i2017-10141-9} {\bibfield  {journal}
  {\bibinfo  {journal} {Riv. Nuovo Cimento}\ }\textbf {\bibinfo {volume}
  {40}},\ \bibinfo {pages} {523} (\bibinfo {year} {2017})}\BibitemShut
  {NoStop}%
\bibitem [{\citenamefont {Nayak}\ \emph {et~al.}(2008)\citenamefont {Nayak},
  \citenamefont {Simon}, \citenamefont {Stern}, \citenamefont {Freedman},\ and\
  \citenamefont {{Das Sarma}}}]{nayak_non-abelian_2008}%
  \BibitemOpen
  \bibfield  {author} {\bibinfo {author} {\bibfnamefont {C.}~\bibnamefont
  {Nayak}}, \bibinfo {author} {\bibfnamefont {S.~H.}\ \bibnamefont {Simon}},
  \bibinfo {author} {\bibfnamefont {A.}~\bibnamefont {Stern}}, \bibinfo
  {author} {\bibfnamefont {M.}~\bibnamefont {Freedman}}, \ and\ \bibinfo
  {author} {\bibfnamefont {S.}~\bibnamefont {{Das Sarma}}},\ }\href@noop {}
  {\bibfield  {journal} {\bibinfo  {journal} {Rev. Mod. Phys.}\ }\textbf
  {\bibinfo {volume} {80}},\ \bibinfo {pages} {1083} (\bibinfo {year}
  {2008})}\BibitemShut {NoStop}%
\bibitem [{\citenamefont {Ivanov}(2001)}]{ivanov01prl268}%
  \BibitemOpen
  \bibfield  {author} {\bibinfo {author} {\bibfnamefont {D.~A.}\ \bibnamefont
  {Ivanov}},\ }\href {\doibase 10.1103/PhysRevLett.86.268} {\bibfield
  {journal} {\bibinfo  {journal} {Phys. Rev. Lett.}\ }\textbf {\bibinfo
  {volume} {86}},\ \bibinfo {pages} {268} (\bibinfo {year} {2001})}\BibitemShut
  {NoStop}%
\bibitem [{\citenamefont {Schnyder}\ \emph {et~al.}(2008)\citenamefont
  {Schnyder}, \citenamefont {Ryu}, \citenamefont {Furusaki},\ and\
  \citenamefont {Ludwig}}]{Schnyder2008}%
  \BibitemOpen
  \bibfield  {author} {\bibinfo {author} {\bibfnamefont {A.~P.}\ \bibnamefont
  {Schnyder}}, \bibinfo {author} {\bibfnamefont {S.}~\bibnamefont {Ryu}},
  \bibinfo {author} {\bibfnamefont {A.}~\bibnamefont {Furusaki}}, \ and\
  \bibinfo {author} {\bibfnamefont {A.~W.~W.}\ \bibnamefont {Ludwig}},\ }\href
  {\doibase 10.1103/PhysRevB.78.195125} {\bibfield  {journal} {\bibinfo
  {journal} {Phys. Rev. B}\ }\textbf {\bibinfo {volume} {78}},\ \bibinfo
  {pages} {195125} (\bibinfo {year} {2008})}\BibitemShut {NoStop}%
\bibitem [{\citenamefont {Kitaev}(2009)}]{Kitaev2009}%
  \BibitemOpen
  \bibfield  {author} {\bibinfo {author} {\bibfnamefont {A.}~\bibnamefont
  {Kitaev}},\ }\href {\doibase 10.1063/1.3149495} {\bibfield  {journal}
  {\bibinfo  {journal} {AIP Conf. Proc.}\ }\textbf {\bibinfo {volume} {1134}},\
  \bibinfo {pages} {22} (\bibinfo {year} {2009})}\BibitemShut {NoStop}%
\bibitem [{\citenamefont {Das}\ \emph {et~al.}(2012)\citenamefont {Das},
  \citenamefont {Ronen}, \citenamefont {Most}, \citenamefont {Oreg},
  \citenamefont {Heiblum},\ and\ \citenamefont {Shtrikman}}]{Anindya2012}%
  \BibitemOpen
  \bibfield  {author} {\bibinfo {author} {\bibfnamefont {A.}~\bibnamefont
  {Das}}, \bibinfo {author} {\bibfnamefont {Y.}~\bibnamefont {Ronen}}, \bibinfo
  {author} {\bibfnamefont {Y.}~\bibnamefont {Most}}, \bibinfo {author}
  {\bibfnamefont {Y.}~\bibnamefont {Oreg}}, \bibinfo {author} {\bibfnamefont
  {M.}~\bibnamefont {Heiblum}}, \ and\ \bibinfo {author} {\bibfnamefont
  {H.}~\bibnamefont {Shtrikman}},\ }\href {\doibase 10.1038/nphys2479}
  {\bibfield  {journal} {\bibinfo  {journal} {Nat. Phys.}\ }\textbf {\bibinfo
  {volume} {8}},\ \bibinfo {pages} {887} (\bibinfo {year} {2012})}\BibitemShut
  {NoStop}%
\bibitem [{\citenamefont {Mourik}\ \emph {et~al.}(2012)\citenamefont {Mourik},
  \citenamefont {Zuo}, \citenamefont {Frolov}, \citenamefont {Plissard},
  \citenamefont {Bakkers},\ and\ \citenamefont {Kouwenhoven}}]{Mourik2012}%
  \BibitemOpen
  \bibfield  {author} {\bibinfo {author} {\bibfnamefont {V.}~\bibnamefont
  {Mourik}}, \bibinfo {author} {\bibfnamefont {K.}~\bibnamefont {Zuo}},
  \bibinfo {author} {\bibfnamefont {S.~M.}\ \bibnamefont {Frolov}}, \bibinfo
  {author} {\bibfnamefont {S.~R.}\ \bibnamefont {Plissard}}, \bibinfo {author}
  {\bibfnamefont {E.~P. A.~M.}\ \bibnamefont {Bakkers}}, \ and\ \bibinfo
  {author} {\bibfnamefont {L.~P.}\ \bibnamefont {Kouwenhoven}},\ }\href
  {\doibase 10.1126/science.1222360} {\bibfield  {journal} {\bibinfo  {journal}
  {Science}\ }\textbf {\bibinfo {volume} {336}},\ \bibinfo {pages} {1003}
  (\bibinfo {year} {2012})}\BibitemShut {NoStop}%
\bibitem [{\citenamefont {Churchill}\ \emph {et~al.}(2013)\citenamefont
  {Churchill}, \citenamefont {Fatemi}, \citenamefont {Grove-Rasmussen},
  \citenamefont {Deng}, \citenamefont {Caroff}, \citenamefont {Xu},\ and\
  \citenamefont {Marcus}}]{Churchil2013}%
  \BibitemOpen
  \bibfield  {author} {\bibinfo {author} {\bibfnamefont {H.~O.~H.}\
  \bibnamefont {Churchill}}, \bibinfo {author} {\bibfnamefont {V.}~\bibnamefont
  {Fatemi}}, \bibinfo {author} {\bibfnamefont {K.}~\bibnamefont
  {Grove-Rasmussen}}, \bibinfo {author} {\bibfnamefont {M.~T.}\ \bibnamefont
  {Deng}}, \bibinfo {author} {\bibfnamefont {P.}~\bibnamefont {Caroff}},
  \bibinfo {author} {\bibfnamefont {H.~Q.}\ \bibnamefont {Xu}}, \ and\ \bibinfo
  {author} {\bibfnamefont {C.~M.}\ \bibnamefont {Marcus}},\ }\href
  {https://link.aps.org/doi/10.1103/PhysRevB.87.241401} {\bibfield  {journal}
  {\bibinfo  {journal} {Phys. Rev. B}\ }\textbf {\bibinfo {volume} {87}},\
  \bibinfo {pages} {241401} (\bibinfo {year} {2013})}\BibitemShut {NoStop}%
\bibitem [{\citenamefont {Nadj-Perge}\ \emph {et~al.}(2014)\citenamefont
  {Nadj-Perge}, \citenamefont {Drozdov}, \citenamefont {Li}, \citenamefont
  {Chen}, \citenamefont {Jeon}, \citenamefont {Seo}, \citenamefont {MacDonald},
  \citenamefont {Bernevig},\ and\ \citenamefont {Yazdani}}]{Yazdani14}%
  \BibitemOpen
  \bibfield  {author} {\bibinfo {author} {\bibfnamefont {S.}~\bibnamefont
  {Nadj-Perge}}, \bibinfo {author} {\bibfnamefont {I.~K.}\ \bibnamefont
  {Drozdov}}, \bibinfo {author} {\bibfnamefont {J.}~\bibnamefont {Li}},
  \bibinfo {author} {\bibfnamefont {H.}~\bibnamefont {Chen}}, \bibinfo {author}
  {\bibfnamefont {S.}~\bibnamefont {Jeon}}, \bibinfo {author} {\bibfnamefont
  {J.}~\bibnamefont {Seo}}, \bibinfo {author} {\bibfnamefont {A.~H.}\
  \bibnamefont {MacDonald}}, \bibinfo {author} {\bibfnamefont {B.~A.}\
  \bibnamefont {Bernevig}}, \ and\ \bibinfo {author} {\bibfnamefont
  {A.}~\bibnamefont {Yazdani}},\ }\href {\doibase 10.1126/science.1259327}
  {\bibfield  {journal} {\bibinfo  {journal} {Science}\ }\textbf {\bibinfo
  {volume} {346}},\ \bibinfo {pages} {602} (\bibinfo {year}
  {2014})}\BibitemShut {NoStop}%
\bibitem [{\citenamefont {L\"othman}\ and\ \citenamefont
  {Black-Schaffer}(2014)}]{Lothman2014}%
  \BibitemOpen
  \bibfield  {author} {\bibinfo {author} {\bibfnamefont {T.}~\bibnamefont
  {L\"othman}}\ and\ \bibinfo {author} {\bibfnamefont {A.~M.}\ \bibnamefont
  {Black-Schaffer}},\ }\href {\doibase 10.1103/PhysRevB.90.224504} {\bibfield
  {journal} {\bibinfo  {journal} {Phys. Rev. B}\ }\textbf {\bibinfo {volume}
  {90}},\ \bibinfo {pages} {224504} (\bibinfo {year} {2014})}\BibitemShut
  {NoStop}%
\bibitem [{\citenamefont {Pawlak}\ \emph {et~al.}(2016)\citenamefont {Pawlak},
  \citenamefont {Kisiel}, \citenamefont {Klinovaja}, \citenamefont {Meier},
  \citenamefont {Kawai}, \citenamefont {Glatzel}, \citenamefont {Loss},\ and\
  \citenamefont {Meyer}}]{Pawlak2015}%
  \BibitemOpen
  \bibfield  {author} {\bibinfo {author} {\bibfnamefont {R.}~\bibnamefont
  {Pawlak}}, \bibinfo {author} {\bibfnamefont {M.}~\bibnamefont {Kisiel}},
  \bibinfo {author} {\bibfnamefont {J.}~\bibnamefont {Klinovaja}}, \bibinfo
  {author} {\bibfnamefont {T.}~\bibnamefont {Meier}}, \bibinfo {author}
  {\bibfnamefont {S.}~\bibnamefont {Kawai}}, \bibinfo {author} {\bibfnamefont
  {T.}~\bibnamefont {Glatzel}}, \bibinfo {author} {\bibfnamefont
  {D.}~\bibnamefont {Loss}}, \ and\ \bibinfo {author} {\bibfnamefont
  {E.}~\bibnamefont {Meyer}},\ }\href {\doibase 10.1038/npjqi.2016.35}
  {\bibfield  {journal} {\bibinfo  {journal} {{n}pj Quantum Inf.}\ }\textbf
  {\bibinfo {volume} {2}},\ \bibinfo {pages} {16035} (\bibinfo {year}
  {2016})}\BibitemShut {NoStop}%
\bibitem [{\citenamefont {Ruby}\ \emph {et~al.}(2017)\citenamefont {Ruby},
  \citenamefont {Heinrich}, \citenamefont {Peng}, \citenamefont {{Von Oppen}},\
  and\ \citenamefont {Franke}}]{Ruby17}%
  \BibitemOpen
  \bibfield  {author} {\bibinfo {author} {\bibfnamefont {M.}~\bibnamefont
  {Ruby}}, \bibinfo {author} {\bibfnamefont {B.~W.}\ \bibnamefont {Heinrich}},
  \bibinfo {author} {\bibfnamefont {Y.}~\bibnamefont {Peng}}, \bibinfo {author}
  {\bibfnamefont {F.}~\bibnamefont {{Von Oppen}}}, \ and\ \bibinfo {author}
  {\bibfnamefont {K.~J.}\ \bibnamefont {Franke}},\ }\href {\doibase
  10.1021/acs.nanolett.7b01728} {\bibfield  {journal} {\bibinfo  {journal}
  {Nano Lett.}\ }\textbf {\bibinfo {volume} {17}},\ \bibinfo {pages} {4473}
  (\bibinfo {year} {2017})}\BibitemShut {NoStop}%
\bibitem [{\citenamefont {M{\'{e}}nard}\ \emph {et~al.}(2017)\citenamefont
  {M{\'{e}}nard}, \citenamefont {Guissart}, \citenamefont {Brun}, \citenamefont
  {Leriche}, \citenamefont {Trif}, \citenamefont {Debontridder}, \citenamefont
  {Demaille}, \citenamefont {Roditchev}, \citenamefont {Simon},\ and\
  \citenamefont {Cren}}]{Menard17}%
  \BibitemOpen
  \bibfield  {author} {\bibinfo {author} {\bibfnamefont {G.~C.}\ \bibnamefont
  {M{\'{e}}nard}}, \bibinfo {author} {\bibfnamefont {S.}~\bibnamefont
  {Guissart}}, \bibinfo {author} {\bibfnamefont {C.}~\bibnamefont {Brun}},
  \bibinfo {author} {\bibfnamefont {R.~T.}\ \bibnamefont {Leriche}}, \bibinfo
  {author} {\bibfnamefont {M.}~\bibnamefont {Trif}}, \bibinfo {author}
  {\bibfnamefont {F.}~\bibnamefont {Debontridder}}, \bibinfo {author}
  {\bibfnamefont {D.}~\bibnamefont {Demaille}}, \bibinfo {author}
  {\bibfnamefont {D.}~\bibnamefont {Roditchev}}, \bibinfo {author}
  {\bibfnamefont {P.}~\bibnamefont {Simon}}, \ and\ \bibinfo {author}
  {\bibfnamefont {T.}~\bibnamefont {Cren}},\ }\href {\doibase
  10.1038/s41467-017-02192-x} {\bibfield  {journal} {\bibinfo  {journal} {Nat.
  Commun.}\ }\textbf {\bibinfo {volume} {8}},\ \bibinfo {pages} {2040}
  (\bibinfo {year} {2017})}\BibitemShut {NoStop}%
\bibitem [{\citenamefont {Kim}\ \emph {et~al.}(2018)\citenamefont {Kim},
  \citenamefont {Palacio-Morales}, \citenamefont {Posske}, \citenamefont
  {Rózsa}, \citenamefont {Palotás}, \citenamefont {Szunyogh}, \citenamefont
  {Thorwart},\ and\ \citenamefont {Wiesendanger}}]{Wiesendanger18}%
  \BibitemOpen
  \bibfield  {author} {\bibinfo {author} {\bibfnamefont {H.}~\bibnamefont
  {Kim}}, \bibinfo {author} {\bibfnamefont {A.}~\bibnamefont
  {Palacio-Morales}}, \bibinfo {author} {\bibfnamefont {T.}~\bibnamefont
  {Posske}}, \bibinfo {author} {\bibfnamefont {L.}~\bibnamefont {Rózsa}},
  \bibinfo {author} {\bibfnamefont {K.}~\bibnamefont {Palotás}}, \bibinfo
  {author} {\bibfnamefont {L.}~\bibnamefont {Szunyogh}}, \bibinfo {author}
  {\bibfnamefont {M.}~\bibnamefont {Thorwart}}, \ and\ \bibinfo {author}
  {\bibfnamefont {R.}~\bibnamefont {Wiesendanger}},\ }\href {\doibase
  10.1126/sciadv.aar5251} {\bibfield  {journal} {\bibinfo  {journal} {Sci.
  Adv.}\ }\textbf {\bibinfo {volume} {4}},\ \bibinfo {pages} {eaar5251}
  (\bibinfo {year} {2018})}\BibitemShut {NoStop}%
\bibitem [{\citenamefont {Palacio-Morales}\ \emph {et~al.}(2019)\citenamefont
  {Palacio-Morales}, \citenamefont {Mascot}, \citenamefont {Cocklin},
  \citenamefont {Kim}, \citenamefont {Rachel}, \citenamefont {Morr},\ and\
  \citenamefont {Wiesendanger}}]{Palacio18}%
  \BibitemOpen
  \bibfield  {author} {\bibinfo {author} {\bibfnamefont {A.}~\bibnamefont
  {Palacio-Morales}}, \bibinfo {author} {\bibfnamefont {E.}~\bibnamefont
  {Mascot}}, \bibinfo {author} {\bibfnamefont {S.}~\bibnamefont {Cocklin}},
  \bibinfo {author} {\bibfnamefont {H.}~\bibnamefont {Kim}}, \bibinfo {author}
  {\bibfnamefont {S.}~\bibnamefont {Rachel}}, \bibinfo {author} {\bibfnamefont
  {D.~K.}\ \bibnamefont {Morr}}, \ and\ \bibinfo {author} {\bibfnamefont
  {R.}~\bibnamefont {Wiesendanger}},\ }\href {\doibase 10.1126/sciadv.aav6600}
  {\bibfield  {journal} {\bibinfo  {journal} {Sci. Adv.}\ }\textbf {\bibinfo
  {volume} {5}},\ \bibinfo {pages} {eaav6600} (\bibinfo {year}
  {2019})}\BibitemShut {NoStop}%
\bibitem [{\citenamefont {Steinbrecher}\ \emph {et~al.}(2018)\citenamefont
  {Steinbrecher}, \citenamefont {Rausch}, \citenamefont {That}, \citenamefont
  {Hermenau}, \citenamefont {Khajetoorians}, \citenamefont {Potthoff},
  \citenamefont {Wiesendanger},\ and\ \citenamefont
  {Wiebe}}]{Steinbrecher2018}%
  \BibitemOpen
  \bibfield  {author} {\bibinfo {author} {\bibfnamefont {M.}~\bibnamefont
  {Steinbrecher}}, \bibinfo {author} {\bibfnamefont {R.}~\bibnamefont
  {Rausch}}, \bibinfo {author} {\bibfnamefont {K.~T.}\ \bibnamefont {That}},
  \bibinfo {author} {\bibfnamefont {J.}~\bibnamefont {Hermenau}}, \bibinfo
  {author} {\bibfnamefont {A.~A.}\ \bibnamefont {Khajetoorians}}, \bibinfo
  {author} {\bibfnamefont {M.}~\bibnamefont {Potthoff}}, \bibinfo {author}
  {\bibfnamefont {R.}~\bibnamefont {Wiesendanger}}, \ and\ \bibinfo {author}
  {\bibfnamefont {J.}~\bibnamefont {Wiebe}},\ }\href {\doibase
  10.1038/s41467-018-05364-5} {\bibfield  {journal} {\bibinfo  {journal} {Nat.
  Commun.}\ }\textbf {\bibinfo {volume} {9}},\ \bibinfo {pages} {2853}
  (\bibinfo {year} {2018})}\BibitemShut {NoStop}%
\bibitem [{\citenamefont {Choi}\ \emph {et~al.}(2019)\citenamefont {Choi},
  \citenamefont {Lorente}, \citenamefont {Wiebe}, \citenamefont {von Bergmann},
  \citenamefont {Otte},\ and\ \citenamefont {Heinrich}}]{Choi19}%
  \BibitemOpen
  \bibfield  {author} {\bibinfo {author} {\bibfnamefont {D.-J.}\ \bibnamefont
  {Choi}}, \bibinfo {author} {\bibfnamefont {N.}~\bibnamefont {Lorente}},
  \bibinfo {author} {\bibfnamefont {J.}~\bibnamefont {Wiebe}}, \bibinfo
  {author} {\bibfnamefont {K.}~\bibnamefont {von Bergmann}}, \bibinfo {author}
  {\bibfnamefont {A.~F.}\ \bibnamefont {Otte}}, \ and\ \bibinfo {author}
  {\bibfnamefont {A.~J.}\ \bibnamefont {Heinrich}},\ }\href {\doibase
  10.1103/RevModPhys.91.041001} {\bibfield  {journal} {\bibinfo  {journal}
  {Rev. Mod. Phys.}\ }\textbf {\bibinfo {volume} {91}},\ \bibinfo {pages}
  {041001} (\bibinfo {year} {2019})}\BibitemShut {NoStop}%
\bibitem [{\citenamefont {Schneider}\ \emph {et~al.}(2021)\citenamefont
  {Schneider}, \citenamefont {Beck}, \citenamefont {Posske}, \citenamefont
  {Crawford}, \citenamefont {Mascot}, \citenamefont {Rachel}, \citenamefont
  {Wiesendanger},\ and\ \citenamefont {Wiebe}}]{schneider_topological_2021}%
  \BibitemOpen
  \bibfield  {author} {\bibinfo {author} {\bibfnamefont {L.}~\bibnamefont
  {Schneider}}, \bibinfo {author} {\bibfnamefont {P.}~\bibnamefont {Beck}},
  \bibinfo {author} {\bibfnamefont {T.}~\bibnamefont {Posske}}, \bibinfo
  {author} {\bibfnamefont {D.}~\bibnamefont {Crawford}}, \bibinfo {author}
  {\bibfnamefont {E.}~\bibnamefont {Mascot}}, \bibinfo {author} {\bibfnamefont
  {S.}~\bibnamefont {Rachel}}, \bibinfo {author} {\bibfnamefont
  {R.}~\bibnamefont {Wiesendanger}}, \ and\ \bibinfo {author} {\bibfnamefont
  {J.}~\bibnamefont {Wiebe}},\ }\href
  {https://www.nature.com/articles/s41567-021-01234-y} {\bibfield  {journal}
  {\bibinfo  {journal} {Nat. Phys.}\ }\textbf {\bibinfo {volume} {17}},\
  \bibinfo {pages} {943} (\bibinfo {year} {2021})}\BibitemShut {NoStop}%
\bibitem [{\citenamefont {Mashkoori}\ and\ \citenamefont
  {Black-Schaffer}(2018)}]{MashkooriDwave}%
  \BibitemOpen
  \bibfield  {author} {\bibinfo {author} {\bibfnamefont {M.}~\bibnamefont
  {Mashkoori}}\ and\ \bibinfo {author} {\bibfnamefont {A.}~\bibnamefont
  {Black-Schaffer}},\ }\href {\doibase 10.1103/PhysRevB.99.024505} {\bibfield
  {journal} {\bibinfo  {journal} {Phys. Rev. B}\ }\textbf {\bibinfo {volume}
  {99}},\ \bibinfo {pages} {24505} (\bibinfo {year} {2018})}\BibitemShut
  {NoStop}%
\bibitem [{\citenamefont {Mashkoori}\ \emph {et~al.}(2020)\citenamefont
  {Mashkoori}, \citenamefont {Pradhan}, \citenamefont {Bj{\"{o}}rnson},
  \citenamefont {Fransson},\ and\ \citenamefont
  {Black-Schaffer}}]{Mashkoori2020}%
  \BibitemOpen
  \bibfield  {author} {\bibinfo {author} {\bibfnamefont {M.}~\bibnamefont
  {Mashkoori}}, \bibinfo {author} {\bibfnamefont {S.}~\bibnamefont {Pradhan}},
  \bibinfo {author} {\bibfnamefont {K.}~\bibnamefont {Bj{\"{o}}rnson}},
  \bibinfo {author} {\bibfnamefont {J.}~\bibnamefont {Fransson}}, \ and\
  \bibinfo {author} {\bibfnamefont {A.~M.}\ \bibnamefont {Black-Schaffer}},\
  }\href {\doibase 10.1103/physrevb.102.104501} {\bibfield  {journal} {\bibinfo
   {journal} {Phys. Rev. B}\ }\textbf {\bibinfo {volume} {102}},\ \bibinfo
  {pages} {104501} (\bibinfo {year} {2020})}\BibitemShut {NoStop}%
\bibitem [{\citenamefont {Fu}\ and\ \citenamefont {Berg}(2010)}]{Fu2010}%
  \BibitemOpen
  \bibfield  {author} {\bibinfo {author} {\bibfnamefont {L.}~\bibnamefont
  {Fu}}\ and\ \bibinfo {author} {\bibfnamefont {E.}~\bibnamefont {Berg}},\
  }\href {\doibase 10.1103/PhysRevLett.105.097001} {\bibfield  {journal}
  {\bibinfo  {journal} {Phys. Rev. Lett.}\ }\textbf {\bibinfo {volume} {105}},\
  \bibinfo {pages} {097001} (\bibinfo {year} {2010})}\BibitemShut {NoStop}%
\bibitem [{\citenamefont {Wray}\ \emph {et~al.}(2010)\citenamefont {Wray},
  \citenamefont {Xu}, \citenamefont {Xia}, \citenamefont {Hor}, \citenamefont
  {Qian}, \citenamefont {Fedorov}, \citenamefont {Lin}, \citenamefont
  {Bansil},\ and\ \citenamefont {Cava}}]{Wray2010}%
  \BibitemOpen
  \bibfield  {author} {\bibinfo {author} {\bibfnamefont {L.~A.}\ \bibnamefont
  {Wray}}, \bibinfo {author} {\bibfnamefont {S.-Y.}\ \bibnamefont {Xu}},
  \bibinfo {author} {\bibfnamefont {Y.}~\bibnamefont {Xia}}, \bibinfo {author}
  {\bibfnamefont {Y.~S.}\ \bibnamefont {Hor}}, \bibinfo {author} {\bibfnamefont
  {D.}~\bibnamefont {Qian}}, \bibinfo {author} {\bibfnamefont {A.~V.}\
  \bibnamefont {Fedorov}}, \bibinfo {author} {\bibfnamefont {H.}~\bibnamefont
  {Lin}}, \bibinfo {author} {\bibfnamefont {A.}~\bibnamefont {Bansil}}, \ and\
  \bibinfo {author} {\bibfnamefont {M.~Z.}\ \bibnamefont {Cava}, \bibfnamefont
  {Robert J.~Hasan}},\ }\href {\doibase 10.1038/nphys1762} {\bibfield
  {journal} {\bibinfo  {journal} {Nat. Phys.}\ }\textbf {\bibinfo {volume}
  {6}},\ \bibinfo {pages} {855} (\bibinfo {year} {2010})}\BibitemShut {NoStop}%
\bibitem [{\citenamefont {Zhang}\ \emph {et~al.}(2013)\citenamefont {Zhang},
  \citenamefont {Kane},\ and\ \citenamefont {Mele}}]{Zhang13}%
  \BibitemOpen
  \bibfield  {author} {\bibinfo {author} {\bibfnamefont {F.}~\bibnamefont
  {Zhang}}, \bibinfo {author} {\bibfnamefont {C.~L.}\ \bibnamefont {Kane}}, \
  and\ \bibinfo {author} {\bibfnamefont {E.~J.}\ \bibnamefont {Mele}},\ }\href
  {\doibase 10.1103/PhysRevLett.111.056402} {\bibfield  {journal} {\bibinfo
  {journal} {Phys. Rev. Lett.}\ }\textbf {\bibinfo {volume} {111}},\ \bibinfo
  {pages} {056402} (\bibinfo {year} {2013})}\BibitemShut {NoStop}%
\bibitem [{\citenamefont {Leijnse}\ and\ \citenamefont
  {Flensberg}(2012)}]{Leijnse12}%
  \BibitemOpen
  \bibfield  {author} {\bibinfo {author} {\bibfnamefont {M.}~\bibnamefont
  {Leijnse}}\ and\ \bibinfo {author} {\bibfnamefont {K.}~\bibnamefont
  {Flensberg}},\ }\href {http://stacks.iop.org/0268-1242/27/i=12/a=124003}
  {\bibfield  {journal} {\bibinfo  {journal} {Semicond. Sci. Technol.}\
  }\textbf {\bibinfo {volume} {27}},\ \bibinfo {pages} {124003} (\bibinfo
  {year} {2012})}\BibitemShut {NoStop}%
\bibitem [{\citenamefont {Beenakker}(2015)}]{BeenakkerRMP}%
  \BibitemOpen
  \bibfield  {author} {\bibinfo {author} {\bibfnamefont {C.~W.~J.}\
  \bibnamefont {Beenakker}},\ }\href {\doibase 10.1103/RevModPhys.87.1037}
  {\bibfield  {journal} {\bibinfo  {journal} {Rev. Mod. Phys.}\ }\textbf
  {\bibinfo {volume} {87}},\ \bibinfo {pages} {1037} (\bibinfo {year}
  {2015})}\BibitemShut {NoStop}%
\bibitem [{\citenamefont {Mascot}\ \emph {et~al.}(2019)\citenamefont {Mascot},
  \citenamefont {Agrahar}, \citenamefont {Rachel},\ and\ \citenamefont
  {Morr}}]{Mascot19}%
  \BibitemOpen
  \bibfield  {author} {\bibinfo {author} {\bibfnamefont {E.}~\bibnamefont
  {Mascot}}, \bibinfo {author} {\bibfnamefont {C.}~\bibnamefont {Agrahar}},
  \bibinfo {author} {\bibfnamefont {S.}~\bibnamefont {Rachel}}, \ and\ \bibinfo
  {author} {\bibfnamefont {D.~K.}\ \bibnamefont {Morr}},\ }\href {\doibase
  10.1103/PhysRevB.100.235102} {\bibfield  {journal} {\bibinfo  {journal}
  {Phys. Rev. B}\ }\textbf {\bibinfo {volume} {100}},\ \bibinfo {pages}
  {235102} (\bibinfo {year} {2019})}\BibitemShut {NoStop}%
\bibitem [{\citenamefont {Crawford}\ \emph {et~al.}(2020)\citenamefont
  {Crawford}, \citenamefont {Mascot}, \citenamefont {Morr},\ and\ \citenamefont
  {Rachel}}]{crawford-20prb174510}%
  \BibitemOpen
  \bibfield  {author} {\bibinfo {author} {\bibfnamefont {D.}~\bibnamefont
  {Crawford}}, \bibinfo {author} {\bibfnamefont {E.}~\bibnamefont {Mascot}},
  \bibinfo {author} {\bibfnamefont {D.~K.}\ \bibnamefont {Morr}}, \ and\
  \bibinfo {author} {\bibfnamefont {S.}~\bibnamefont {Rachel}},\ }\href
  {\doibase 10.1103/PhysRevB.101.174510} {\bibfield  {journal} {\bibinfo
  {journal} {Phys. Rev. B}\ }\textbf {\bibinfo {volume} {101}},\ \bibinfo
  {pages} {174510} (\bibinfo {year} {2020})}\BibitemShut {NoStop}%
\bibitem [{\citenamefont {Mashkoori}\ \emph {et~al.}(2019)\citenamefont
  {Mashkoori}, \citenamefont {Moghaddam}, \citenamefont {Hajibabaee},
  \citenamefont {Black-Schaffer},\ and\ \citenamefont
  {Parhizgar}}]{Mashkoori2018}%
  \BibitemOpen
  \bibfield  {author} {\bibinfo {author} {\bibfnamefont {M.}~\bibnamefont
  {Mashkoori}}, \bibinfo {author} {\bibfnamefont {A.~G.}\ \bibnamefont
  {Moghaddam}}, \bibinfo {author} {\bibfnamefont {M.~H.}\ \bibnamefont
  {Hajibabaee}}, \bibinfo {author} {\bibfnamefont {A.~M.}\ \bibnamefont
  {Black-Schaffer}}, \ and\ \bibinfo {author} {\bibfnamefont {F.}~\bibnamefont
  {Parhizgar}},\ }\href {\doibase 10.1103/PhysRevB.99.014508} {\bibfield
  {journal} {\bibinfo  {journal} {Phys. Rev. B}\ }\textbf {\bibinfo {volume}
  {99}},\ \bibinfo {pages} {14508} (\bibinfo {year} {2019})}\BibitemShut
  {NoStop}%
\bibitem [{\citenamefont {Fulga}\ \emph {et~al.}(2012)\citenamefont {Fulga},
  \citenamefont {Akhmerov}, \citenamefont {Tworzyd\l{}o}, \citenamefont
  {B\'eri},\ and\ \citenamefont {Beenakker}}]{Fulga_2012}%
  \BibitemOpen
  \bibfield  {author} {\bibinfo {author} {\bibfnamefont {I.~C.}\ \bibnamefont
  {Fulga}}, \bibinfo {author} {\bibfnamefont {A.~R.}\ \bibnamefont {Akhmerov}},
  \bibinfo {author} {\bibfnamefont {J.}~\bibnamefont {Tworzyd\l{}o}}, \bibinfo
  {author} {\bibfnamefont {B.}~\bibnamefont {B\'eri}}, \ and\ \bibinfo {author}
  {\bibfnamefont {C.~W.~J.}\ \bibnamefont {Beenakker}},\ }\href {\doibase
  10.1103/PhysRevB.86.054505} {\bibfield  {journal} {\bibinfo  {journal} {Phys.
  Rev. B}\ }\textbf {\bibinfo {volume} {86}},\ \bibinfo {pages} {054505}
  (\bibinfo {year} {2012})}\BibitemShut {NoStop}%
\bibitem [{\citenamefont {Diez}\ \emph {et~al.}(2014)\citenamefont {Diez},
  \citenamefont {Fulga}, \citenamefont {Pikulin}, \citenamefont {Tworzydło},\
  and\ \citenamefont {Beenakker}}]{Diez_2014}%
  \BibitemOpen
  \bibfield  {author} {\bibinfo {author} {\bibfnamefont {M.}~\bibnamefont
  {Diez}}, \bibinfo {author} {\bibfnamefont {I.~C.}\ \bibnamefont {Fulga}},
  \bibinfo {author} {\bibfnamefont {D.~I.}\ \bibnamefont {Pikulin}}, \bibinfo
  {author} {\bibfnamefont {J.}~\bibnamefont {Tworzydło}}, \ and\ \bibinfo
  {author} {\bibfnamefont {C.~W.~J.}\ \bibnamefont {Beenakker}},\ }\href
  {\doibase 10.1088/1367-2630/16/6/063049} {\bibfield  {journal} {\bibinfo
  {journal} {New Journal of Physics}\ }\textbf {\bibinfo {volume} {16}},\
  \bibinfo {pages} {063049} (\bibinfo {year} {2014})}\BibitemShut {NoStop}%
\bibitem [{\citenamefont {Qi}\ \emph {et~al.}(2010)\citenamefont {Qi},
  \citenamefont {Hughes},\ and\ \citenamefont {Zhang}}]{Qi2010}%
  \BibitemOpen
  \bibfield  {author} {\bibinfo {author} {\bibfnamefont {X.-L.}\ \bibnamefont
  {Qi}}, \bibinfo {author} {\bibfnamefont {T.~L.}\ \bibnamefont {Hughes}}, \
  and\ \bibinfo {author} {\bibfnamefont {S.-C.}\ \bibnamefont {Zhang}},\ }\href
  {\doibase 10.1103/PhysRevB.81.134508} {\bibfield  {journal} {\bibinfo
  {journal} {Phys. Rev. B}\ }\textbf {\bibinfo {volume} {81}},\ \bibinfo
  {pages} {134508} (\bibinfo {year} {2010})}\BibitemShut {NoStop}%
\bibitem [{\citenamefont {Teo}\ and\ \citenamefont {Kane}(2010)}]{Teo2010}%
  \BibitemOpen
  \bibfield  {author} {\bibinfo {author} {\bibfnamefont {J.~C.~Y.}\
  \bibnamefont {Teo}}\ and\ \bibinfo {author} {\bibfnamefont {C.~L.}\
  \bibnamefont {Kane}},\ }\href {\doibase 10.1103/PhysRevB.82.115120}
  {\bibfield  {journal} {\bibinfo  {journal} {Phys. Rev. B}\ }\textbf {\bibinfo
  {volume} {82}},\ \bibinfo {pages} {115120} (\bibinfo {year}
  {2010})}\BibitemShut {NoStop}%
\bibitem [{\citenamefont {Bj\"ornson}\ and\ \citenamefont
  {Black-Schaffer}(2013)}]{Bjornson2013}%
  \BibitemOpen
  \bibfield  {author} {\bibinfo {author} {\bibfnamefont {K.}~\bibnamefont
  {Bj\"ornson}}\ and\ \bibinfo {author} {\bibfnamefont {A.~M.}\ \bibnamefont
  {Black-Schaffer}},\ }\href {\doibase 10.1103/PhysRevB.88.024501} {\bibfield
  {journal} {\bibinfo  {journal} {Phys. Rev. B}\ }\textbf {\bibinfo {volume}
  {88}},\ \bibinfo {pages} {024501} (\bibinfo {year} {2013})}\BibitemShut
  {NoStop}%
\bibitem [{\citenamefont {Bj{\"{o}}rnson}\ \emph {et~al.}(2015)\citenamefont
  {Bj{\"{o}}rnson}, \citenamefont {Pershoguba}, \citenamefont {Balatsky},\ and\
  \citenamefont {Black-Schaffer}}]{Bjornson2015}%
  \BibitemOpen
  \bibfield  {author} {\bibinfo {author} {\bibfnamefont {K.}~\bibnamefont
  {Bj{\"{o}}rnson}}, \bibinfo {author} {\bibfnamefont {S.~S.}\ \bibnamefont
  {Pershoguba}}, \bibinfo {author} {\bibfnamefont {A.~V.}\ \bibnamefont
  {Balatsky}}, \ and\ \bibinfo {author} {\bibfnamefont {A.~M.}\ \bibnamefont
  {Black-Schaffer}},\ }\href {\doibase 10.1103/PhysRevB.92.214501} {\bibfield
  {journal} {\bibinfo  {journal} {Phys. Rev. B}\ }\textbf {\bibinfo {volume}
  {92}},\ \bibinfo {pages} {214501} (\bibinfo {year} {2015})}\BibitemShut
  {NoStop}%
\bibitem [{\citenamefont {Bj\"ornson}\ and\ \citenamefont
  {Black-Schaffer}(2016)}]{Bjornson2016}%
  \BibitemOpen
  \bibfield  {author} {\bibinfo {author} {\bibfnamefont {K.}~\bibnamefont
  {Bj\"ornson}}\ and\ \bibinfo {author} {\bibfnamefont {A.~M.}\ \bibnamefont
  {Black-Schaffer}},\ }\href {\doibase 10.1103/PhysRevB.94.100501} {\bibfield
  {journal} {\bibinfo  {journal} {Phys. Rev. B}\ }\textbf {\bibinfo {volume}
  {94}},\ \bibinfo {pages} {100501} (\bibinfo {year} {2016})}\BibitemShut
  {NoStop}%
\bibitem [{\citenamefont {Björnson}(2019)}]{KristoferTBTK}%
  \BibitemOpen
  \bibfield  {author} {\bibinfo {author} {\bibfnamefont {K.}~\bibnamefont
  {Björnson}},\ }\href {\doibase https://doi.org/10.1016/j.softx.2019.02.005}
  {\bibfield  {journal} {\bibinfo  {journal} {SoftwareX}\ }\textbf {\bibinfo
  {volume} {9}},\ \bibinfo {pages} {205} (\bibinfo {year} {2019})}\BibitemShut
  {NoStop}%
\bibitem [{\citenamefont {Anderson}(1959)}]{Anderson59}%
  \BibitemOpen
  \bibfield  {author} {\bibinfo {author} {\bibfnamefont {P.}~\bibnamefont
  {Anderson}},\ }\href {\doibase https://doi.org/10.1016/0022-3697(59)90036-8}
  {\bibfield  {journal} {\bibinfo  {journal} {J. Phys. Chem. Solids}\ }\textbf
  {\bibinfo {volume} {11}},\ \bibinfo {pages} {26} (\bibinfo {year}
  {1959})}\BibitemShut {NoStop}%
\bibitem [{\citenamefont {Black-Schaffer}(2011)}]{BlackSchaffer11}%
  \BibitemOpen
  \bibfield  {author} {\bibinfo {author} {\bibfnamefont {A.~M.}\ \bibnamefont
  {Black-Schaffer}},\ }\href {\doibase 10.1103/PhysRevB.83.060504} {\bibfield
  {journal} {\bibinfo  {journal} {Phys. Rev. B}\ }\textbf {\bibinfo {volume}
  {83}},\ \bibinfo {pages} {060504} (\bibinfo {year} {2011})}\BibitemShut
  {NoStop}%
\bibitem [{\citenamefont {Borchmann}\ \emph {et~al.}(2016)\citenamefont
  {Borchmann}, \citenamefont {Farrell},\ and\ \citenamefont
  {Pereg-Barnea}}]{Borchmann16}%
  \BibitemOpen
  \bibfield  {author} {\bibinfo {author} {\bibfnamefont {J.}~\bibnamefont
  {Borchmann}}, \bibinfo {author} {\bibfnamefont {A.}~\bibnamefont {Farrell}},
  \ and\ \bibinfo {author} {\bibfnamefont {T.}~\bibnamefont {Pereg-Barnea}},\
  }\href {\doibase 10.1103/PhysRevB.93.125133} {\bibfield  {journal} {\bibinfo
  {journal} {Phys. Rev. B}\ }\textbf {\bibinfo {volume} {93}},\ \bibinfo
  {pages} {125133} (\bibinfo {year} {2016})}\BibitemShut {NoStop}%
\end{thebibliography}%
\end{document}